\documentstyle[11pt,menu97,epsfig]{article}

\begin{document}
\setlength{\baselineskip}{2.6ex}

\title{ What Do We Know about Low--Energy $\bar{K}N$ Interactions?\\
Need and Possibilities of New Experiments at DA$\Phi$NE.\thanks{Research 
supported by the E.E.C. Human Capital and Mobility Program under contract 
No. CHRX--CT92--0026.}}

\author{Paolo M. Gensini, Rafael Hurtado\thanks{On leave of absence from 
Centro Internacional de F\'\i{s}ica, Bogot\'a, Colombia, under a grant by 
Colciencias. This work is part of his Ph.D. Thesis.}\\
{\em Dip. Fisica, Univ. Perugia, and Sez. I.N.F.N., Perugia, Italy}\\
and\\ Galileo Violini\\
{\em Dip. Fisica, Univ. Calabria, and Gr. Coll. I.N.F.N., Cosenza, Italy\\
and\\
Univ. de El Salvador, San Salvador}}

\maketitle

\begin{abstract}
\setlength{\baselineskip}{2.6ex}
We present results from a dispersion--relation investigation on a 
coupled--channel analysis of $\bar{K}N$--initiated processes in the low-- 
and intermediate--energy region. The analysis demonstrates both the 
effectiveness of these relations to constrain the parameters of the 
unmeasurable ${\pi}Y$ channels and the need for better data in the 
low--energy region, such as could be provided at a $\phi$--factory.
\end{abstract}

\setlength{\baselineskip}{2.6ex}

\section*{1. INTRODUCTION.}

In this talk we shall discuss the quality of our knowledge of the 
parameters which describe $\bar{K}N$ low--energy interactions. The 
forthcoming beginning of operations at DA$\Phi$NE provides an 
interesting possibility for improving this knowledge\cite{1}, but a 
motivation for further study has also emerged from our analysis of 
the ${\pi}Y$ channels' parameters presented at the recent MENU '97 
Symposium in Vancouver\cite{2}.

\section*{2. GENERALITIES. }

It is customary to analyse the low--energy $\bar{K}N$ interactions 
using a multichannel K--matrix (or, equivalently, M--matrix, with ${\bf M} 
= {\bf K}^{-1}$) formalism\cite{3}
\begin{equation}
{\bf Q}^\ell({\bf T}_I^\ell){\bf Q}^\ell=({\bf K}_I^\ell)^{-1}-i
{\bf Q}^{(2\ell+1)}\ \ ;
\label{1}
\end{equation}
in this equation ${\bf K}_I^\ell$ is a real Hermitian matrix, possibly 
depending on energy, and ${\bf Q}$ is the diagonal matrix of the c.m. 
momenta $q_i$ for the different $S=-1$ channels. $I$ indicates the 
isospin and $\ell$ the orbital angular momentum of the partial wave 
considered. Neglecting the $\Lambda\pi\pi$ and $\Sigma\pi\pi$ channels, 
as done at these low energies for limited statistics experiments, 
each partial wave of the $\bar{K}N$ 
interactions is described by a matrix of dimensions $2\times2$ for $I=0$ 
and $3\times3$ for $I=1$. Note that in this isospin decomposition the 
$\bar{K}^0N$ channels are considered together with the $K^-N$ ones: this 
neglects mass difference effects which will be important at DA$\Phi$NE, 
since the charge--exchange threshold in $K^-p$ scattering is at a 
lab. momentum $k_L=$ 90 MeV/c. This formalism has been used to analyse 
data from experiments carried out between threshold and a lab. 
momentum $k_L\simeq$ 500 MeV/c, and 
for this purpose S, P and D waves have been considered\cite{4}.

The matrix elements required for an analysis including up to $J=3/2$ are 
36, and if one takes into account their energy dependence, this increases 
further the number of parameters: for instance, in ref. [4a] they are 44.

This problem has been circumvented either by analysing the data using 
only S waves (with only 9 matrix elements)\cite{5}, and reducing 
consequently the highest laboratory momentum considered, or by introducing 
some theoretical constraints on the parameters\cite{6}.

As we shall show, the parametrization of ref. [4a], which includes all 
waves up to $J=3/2$, badly violates obvious consistency conditions coming 
from dispersion relations. This strongly suggests the need for new analyses 
which could take advantage of the much better data DA$\Phi$NE is expected 
to make available, and which could incorporate adequate dispersion relation 
constraints.

We recall that the ${\pi}N$ interaction at low energy, whose analyses make 
a heavy use of such constraints, is very well known, even if some 
uncertainty, of the order of a few percent, still exists on the ${\pi}NN$ 
coupling constant\cite{7}. Our hope is that in a few years the joint work 
of experimentalists and theorists might lead, if not to a comparable 
quality in the knowledge of the $\bar{K}N$ interaction, at least 
to the elimination of the difficulties we are going to present here. The 
reasons for putting so much hope in DA$\Phi$NE are well known\cite{8}: this 
machine will produce kaon pairs in quantities such as to lead to statistics 
orders of magnitude better than those of the experiments analysed in the 
past, and with an excellent momentum resolution.

Even if the main scientific goals of DA$\Phi$NE are direct CP violation in 
$K_{L,S}$ decays, rare decays and bounds on CPT violation, the machine 
offers unique possibilities for measurements of angular distributions and 
polarization of final--state hyperons in $\bar{K}N$ interactions, provided 
a strong scientific case for such experiments exists.

\section*{3. OUTLINE OF THE DISPERSIVE CALCULATIONS. }

The analysis of ref. [4a], of about thirty years ago, is the only one which 
includes up to D waves and extends down to the $\bar{K}N$ threshold, with a 
reasonable behaviour in the unphysical region below that threshold. We shall 
discuss here the inconsistencies put in evidence for that analysis by 
dispersion relations. We do not perform a similar analysis of parametrizations 
which only use S waves, because (i) it is obvious that their region of 
validity is significantly smaller, (ii) our method is particularly sensitive 
to P waves (and we anticipate that they will appear to be very poorly known), 
and (iii) one expects that the decuplet resonance $\Sigma(1385)$ must play a 
significant role, as its non--strange partner $\Delta(1232)$ does in the 
${\pi}N$ case.

Our consistency test is based on the evaluation of the $PBB$ couplings. We 
recall that flavour SU(3) symmetry provides definite predictions for these 
quantities\cite{9} in terms of one of them, $G_{{\pi}^-pn}$, and of a 
parameter $\alpha=F/(F+D)$ which expresses the weight of the antisymmetric 
octet in the direct product $8\times8$.

The $PBB$ coupling constants are usually evaluated by forward dispersion 
relations, and in the case of the $S=-1$ sector a slightly different 
situation occurs for ${\pi}YY'$ couplings with respect to the $KYN$ ones. 
In the case of $KYN$ couplings, since $K^{\pm}N$ total cross sections are 
known up to very high energies, one can write once--subtracted dispersion 
relations for the 
spin--averaged forward amplitude $C(\omega)$, since its imaginary part is 
directly proportional to $\sigma_{tot}$ via the optical theorem. This 
amplitude turns out to be S--wave dominated at very low energies, and this 
has made it possible to determine these couplings from low--energy 
analyses which only included the S waves\cite{10}. Instead, in the case 
of the $\pi\Sigma\Sigma$ and $\pi\Lambda\Sigma$ couplings, the amplitudes 
can only be estimated in the region of validity of the multichannel 
formalism, and the previous application of the optical theorem is clearly 
of no use. Therefore one must use rapidly convergent dispersion relations, 
and this leads to choose the amplitude $B(\omega)/\omega$, even under 
crossing between $s$ and $u$ channels, for elastic $\pi\Lambda$ and 
$\pi\Sigma$ scattering. It has to be noted that these amplitudes, beside 
being rapidly decreasing as $\omega\to\infty$, exhibit a strong dependence 
on the P waves close to threshold.

We also calculated the product of these couplings from the $\pi\Lambda
\to\pi\Sigma$ reaction, using this time the amplitude 
$C(\omega)/\omega$: this channel has been considered here for the first 
time, while the two elastic processes were already studied by other 
authors\cite{11}. However, the novelty of our calculation is not only 
in this part, and in the update of the inputs for the higher--energy 
ranges of the dispersive integrals, but also in the fact that for the 
first time we analysed the stability of the results with respect to 
the energy at which the dispersion relations were evaluated.

This provides an important consistency check of the parametrization of the 
$S=-1$ meson--barion interactions with fixed--$t$ analyticity: we found 
indeed that ${\pi}Y$ amplitudes showed marked inconsistencies and, in order 
to investigate their sources, we analysed separately the contributions from 
each partial wave in the parametrization of the low--energy region, and in 
particular their dependences on the energy at which the dispersions 
relations were evaluated. Such a separation is useful in building a 
priority scale for future experiments to be performed, for instance, at 
DA$\Phi$NE.

\section*{4. DESCRIPTION OF THE RESULTS. }

For what concerns our calculations of the ${\pi}YY'$ couplings, here we shall 
rapidly summarise the results of ref. [2], to which we also refer for the 
details on the evaluation of the dispersive integrals.

In the case of $\pi\Lambda$ elastic scattering we found that the value for 
$G^2_{\pi\Lambda\Sigma}/4\pi$ evaluated at the $\bar{K}N$ threshold was in 
agreement with the similar calculation by Chang and Meiere\cite{11}. 
However, as we varied the energy at which the dispersion relation was 
evaluated, we found significant variations in the coupling as well, almost 
entirely due to the P$_{11}$ wave, the remaining contributions having a 
smooth energy dependence, and accounting at the $\bar{K}N$ threshold for 
about 28.5 \% of the value of $G^2_{\pi\Lambda\Sigma}/4\pi$.

In the case of $\pi\Sigma$ elastic scattering, for which one has to use 
only the (crossing--even) combination of isospin amplitudes $2B_1-B_0$, in 
order to eliminate the unknown amplitude $B_2$ in both the $s$ and $u$ 
channels, both $\Lambda$ and $\Sigma$ poles are present in the Born term: 
thus a reasonable approach to the determination of $G^2_{\pi\Sigma\Sigma}
/4\pi$ is to combine the dispersion relation for this channel with the 
one for the $\pi\Lambda$ channel (calculated at the same c.m. energy), 
so that the $\Lambda$--pole term in the first relation is exactly canceled 
by the $\Sigma$--pole one in the second (this was done as well by Chan and 
Meiere\cite{11}, although only at the $\bar{K}N$ threshold). Our results 
for the $\pi\Sigma\Sigma$ coupling agree with Chan--Meiere's at the 
$\bar{K}N$ threshold, but away from it a complicated energy dependence 
appears, which in part can be ascribed both to the poor matching of Kim's 
analysis to the input used for the intermediate--energy range of the 
integrals (at the upper end of the energy interval covered by our 
calculations), and (at the lower end of that interval) to the $\Lambda$ 
pole falling on the cut between the $\pi\Lambda$ and $\pi\Sigma$ 
thresholds, so that the M--matrices can not describe correctly the real 
part of the amplitude in this energy region. Between these two extremes, 
the coupling shows a hump close to the $\bar{K}N$ threshold, which is 
again, as in the previous case, mostly due to the P$_{11}$ wave.

Finally, when the $\pi\Lambda\to\pi\Sigma$ reaction is analysed, one 
finds several inconsistencies: first, the product $G_{\pi\Lambda\Sigma} 
G_{\pi\Sigma\Sigma}/4\pi$ has a value close to zero at the $\bar{K}N$ 
threshold, in clear contrast with the values found by us and by ref. [11] 
for the separate couplings at the same energy. Second, the product shows 
a marked dependence on the energy, going rapidly from positive to negative 
values: again, this seems to be mostly due to the P$_{11}$ wave, whose 
contribution is negative over most of the range and rapidly decreasing.

However, if the $J=1/2$ P waves are dropped from all three calculations, 
not only the couplings obtained show much smaller variations in the 
central part of the energy range covered by our calculations, but also 
the average values obtained from the three channels are rather consistent 
with each other (and, incidentally, also with the expectations from 
flavour SU(3) symmetry and the known value of $G^2_{{\pi}NN}/4\pi$). This 
seems to suggest that these waves were very poorly determined in Kim's 
analysis, at least as far as their matrix elements in the ${\pi}Y$ 
channels are concerned.

We have therefore analysed also the $\bar{K}N$ elastic scattering $B$ 
amplitudes by the same approach. It is convenient to recall again here 
that the dispersive evaluations of the $KYN$ couplings have usually 
involved the amplitudes $C(\omega)$, mainly sensitive to the S waves 
close to threshold. Nevertheless, an old paper\cite{12} had indeed found 
an energy dependence in $G^2_{KN\Lambda}$, which perhaps could have had 
something to do with the problems we have just pointed out with Kim's 
$J=1/2$ P waves.

The calculation of $G^2_{K\Lambda{N}}/4\pi$ and $G^2_{K\Sigma{N}}/4\pi$ was 
performed using the pure $I=0$ and 1 isospin combinations of $\bar{K}N$ 
$B$ amplitudes, using for the crossed $KN$ channels the VPI phase shift 
analysis\cite{13}. We calculated a dispersion relation for the difference 
${\rm Re}\bar{B}(\omega_1)-{\rm Re}B(\omega_2)$, where $\bar{B}$, $B$ are 
the $\bar{K}N$, $KN$ $B$ amplitudes, and $\omega_2$ was kept fixed while 
$\omega_1$ varied in the energy range covered by Kim's parametrization 
(unphysical range included). Calculations were carried out for four values 
of $\omega_2$, corresponding to $KN$ c.m. energies of 1,468, 1,500, 1,530 
and 1,560 MeV: the results were very similar, so that we only present here 
those referring to the first value, i.e. to $\omega_2$ = 548 MeV.

\parbox{16.5cm}{
\begin{center}
\epsfig{figure=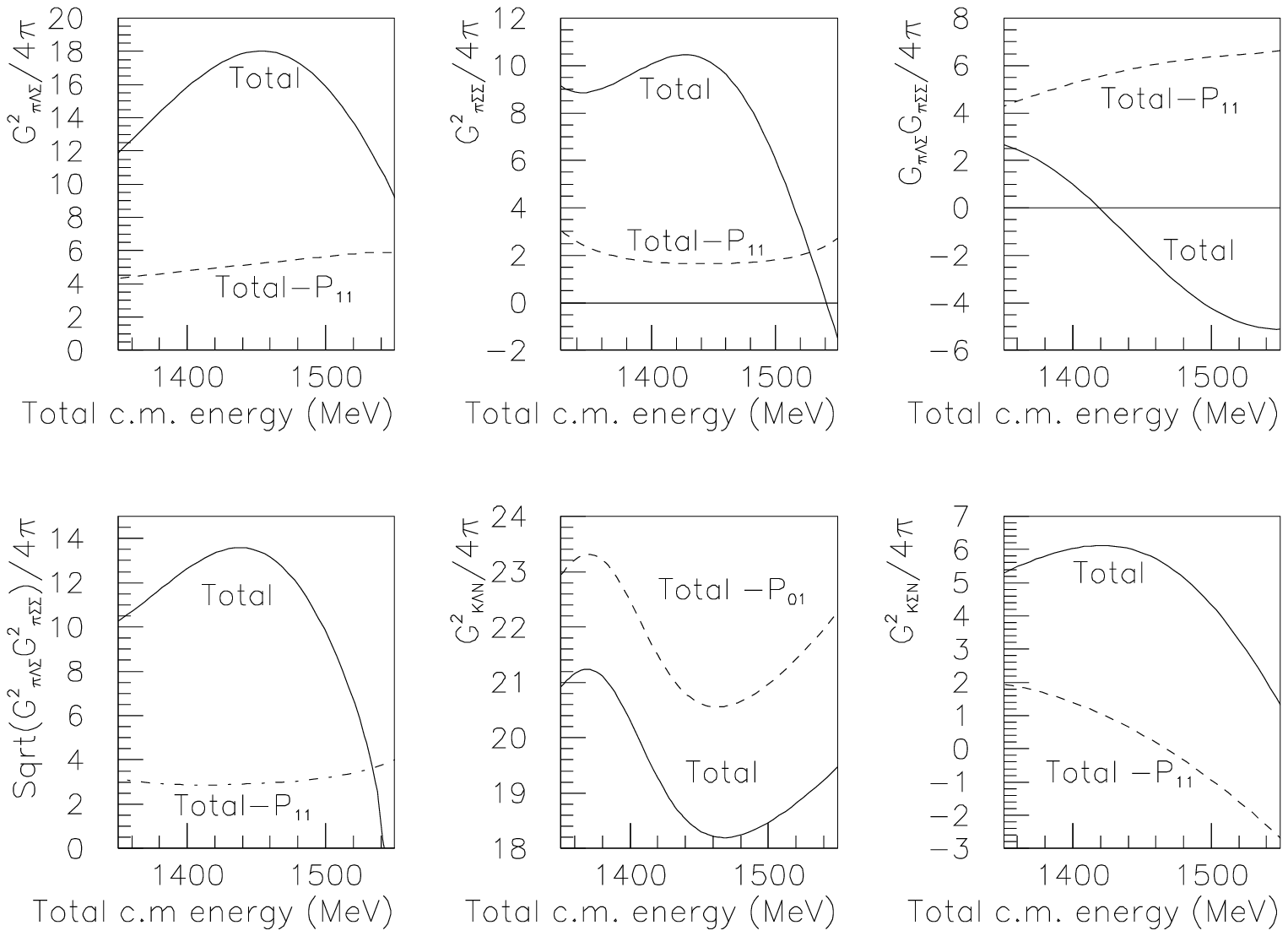,width=16cm,height=12cm}
\end{center}}

\parbox{15.5cm}{
\begin{center}
{\noindent
\parbox{13cm}
{\small \setlength{\baselineskip}{2.6ex} Fig.~1. Results on the couplings 
to the $\Sigma$ pole in, respectively, (first row, from left to right) 
$\pi\Lambda\to\pi\Lambda$, $\pi\Sigma\to\pi\Sigma$ and $\pi\Lambda\to\pi
\Sigma$, to be compared (second row) with the product of the 
first two, and results for the couplings to the $\Lambda$ and $\Sigma$ 
poles in the $I=0$ and 1 channels of $\bar{K}N$ scattering.}}
\end{center}}
\vspace*{1cm}

The $K\Lambda{N}$ coupling constant $G_{K\Lambda{N}}^2/4\pi$ exhibits a 
bump--like structure for a $\bar{K}N$ c.m. energy of about 1,300 through 
1,450 MeV, and a general trend to increase with c.m. energy over the 
whole energy range analysed by us, oscillating around a value of 19.5 in 
the central part of this range. The bump is coming from the contribution 
of the D$_{03}$ wave, while the average slope comes mostly from the 
$u$--channel $KN$ contributions. No significant effect can be attributed 
to the P$_{01}$ wave contribution.

In the case of the $K\Sigma{N}$ coupling, $G_{K\Sigma{N}}^2/4\pi$ exhibits 
a dependence on the $\bar{K}N$ c.m. energy peaking around threshold, which 
is almost completely eliminated when the contributions from the P$_{11}$ 
wave is dropped out.

From these calculations it transpares that the only parametrisation of the 
$S=-1$, low--energy $PB$ scattering, which includes waves up to $J=3/2$, 
and appears to give a reliable description of the structure of the 
$\pi\Lambda$ and $\pi\Sigma$ channels in the $\bar{K}N$ unphysical region, 
is inconsistent with dispersion relations, and that also the VPI 
partial--wave analysis of $KN$ scattering\cite{13} seems to meet some 
difficulties, at least for the combination $B_1-B_0=2B_{K^+p}-B_{K^+n}$. 
However, if one considers the many measurements possible in low--energy 
$\bar{K}N$--initiated reactions, and the few ones actually performed in 
all previous experiments, it is possible to hold the optimistic wiev that 
with DA$\Phi$NE the situation might significantly improve. Before 
discussing this in some detail, let us recall that the use of more 
sophisticated tools of analysis, like those proposed in ref. [1c], can 
also shed more light on this subject.

Going to the experimental perspectives, it has to be stressed that the 
parameters of the K--matrices are determined at the lowest energies from 
very few sources, measured usually a rather long time ago: they are the 
{\em integrated} cross sections for the reactions $K^-p\to{K^-}p$, 
$K^-p\to{K^0_S}n$, $K^-p\to\pi^\pm\Sigma^\mp$, and (but without resolving 
the channels) $K^-p\to\Lambda$ + neutrals.

\section*{5. FUTURE PERSPECTIVES. }

DA$\Phi$NE will make possible to analyse all these reactions and, with 
the calorimetric techniques employed by KLOE, to resolve all charged and 
neutral channels (and detecting the $\gamma$ from $\Sigma^0$ decay to 
separate $\pi^0(\pi^0)\Sigma^0$ from $\pi^0(\pi^0)\Lambda$), not only 
measuring the rates but also the differential distributions, at least to 
the statistical level to extract their Lagrange coefficients $L_1$ and 
$L_2$. The extemely good efficiency and spatial resolution of a KLOE--like 
apparatus, of much smaller dimensions than its ``big brother'' due to the 
much shorter decay length of the $K^\pm$ with respect to that of the 
$K^0_L$, will also allow measurements of the polarizations of the 
final--state $\Lambda$ and $\Sigma^+$ hyperons from the angular 
distributions of their decay products. The same measurements will also be 
possible for all $K^0_Lp$--initiated processes, and, replacing hydrogen 
with deuterium, for $K^0_Ln$-- and $K^-n$--initiated ones as well. 
Already this amount of experimental information at laboratory momenta 
from about 110 down to about 90 MeV/c (the detector itself will act as a 
``moderator'' for $K^\pm$) should lead to a much better knowledge of the 
K--matrices and to more reliable calculations of the quantities related 
to them, such as the $PBB$ coupling constants for the $S=-1$ sector and 
the zero--energy values of the crossing--even amplitudes $C(\omega,t)$, 
which can lead to an estimate of the $KN$ $\sigma$--terms\cite{14}. 
Other interesting experiments will already be possible with existing 
detectors: KLOE\cite{15} will surely be able to register all interactions 
of both $K^\pm$ and $K^0_L$ with the $^4$He filling its wire chamber, 
interactions never observed before at such low lab. momenta, DEAR\cite{16} 
will measure the K lines of kaonic hydrogen (and deuterium) giving 
independent information on the $\bar{K}N$ S--wave scattering lengths 
(with CCDs covering much lower $\gamma$--ray energies they could also 
think about investigating the P waves through the study of the L lines as 
well), and FINUDA\cite{17}, though starting with a much narrower 
scope than KLOE, will anyway be able to make some high quality 
measurements, in particular of the $K^0_Lp$ charge--exchange processes 
taking place in the hydrogen of its plastic scintillators\cite{18}.

We hope with the present discussion to have offered at least part of the 
background we were referring to for the motivation of new, perhaps less 
fashionable than others, but nonetheless still very interesting experiments, 
several of which, despite the folklore about low--energy physics having 
been already adequately explored in the past, were never done before, and 
quite certainly not accomplishable anywhere else but at a $\phi$--factory.

\bibliographystyle{unsrt}

\end{document}